\documentclass[fleqn,11pt]{article}

\usepackage{amsfonts,amsmath,amssymb}
\usepackage{latexsym}

\usepackage{amsfonts,amssymb,cite}
\usepackage{graphicx}

\def\noi{\noindent}
\def\jnumber#1#2{\thispagestyle{empty} \noi\unitlength=1mm
    	\begin{picture}(178,10)
            \put(177,15){\llap{\large\it Grav. Cosmol. No.\,#1, #2}}
                    \end{picture}}

\newcommand{\Title}[1]{\noi {{\Large\bf #1}}\\[1ex]}

\def\Aunames#1{\noi{\bf #1}}
\def\au#1{${}^{#1}$}
\def\Addresses#1{\medskip\noi \protect
	\begin{description}\itemsep -3pt {\it #1} \end{description}}
\def\adr#1#2{\item[${}^{#1}$]{\it #2}}

\newcommand{\Abstract}[1]{\vskip 2mm \begin{center}
        \parbox{16.4cm}{\small\noi #1} \end{center}\medskip}

\def\email#1#2{\footnotetext[#1]{e-mail: #2}\addtocounter{footnote}{1}}

\def\nqq{\hspace*{-2em}}

\def\inch{\hspace*{1in}}

\usepackage{color}

\def\Acknow#1{\subsection*{Acknowledgments} #1}
\def\Funding#1{\subsection*{Funding} #1}

\def\Jl#1#2{#1 {\bf #2},\ }

\def\ApJ#1 {\Jl{Astroph. J.}{#1}}
\def\CQG#1 {\Jl{Class. Quantum Grav.}{#1}}
\def\DAN#1 {\Jl{Dokl. AN SSSR}{#1}}
\def\GC#1 {\Jl{Grav. Cosmol.}{#1}}
\def\GRG#1 {\Jl{Gen. Rel. Grav.}{#1}}
\def\IJMPD#1 {\Jl{Int. J. Mod. Phys. D}{#1}}
\def\JETF#1 {\Jl{Zh. Eksp. Teor. Fiz.}{#1}}
\def\JETP#1 {\Jl{Sov. Phys. JETP}{#1}}
\def\JHEP#1 {\Jl{JHEP}{#1}}
\def\JMP#1 {\Jl{J. Math. Phys.}{#1}}
\def\NPB#1 {\Jl{Nucl. Phys. B}{#1}}
\def\NP#1 {\Jl{Nucl. Phys.}{#1}}
\def\PLA#1 {\Jl{Phys. Lett. A}{#1}}
\def\PLB#1 {\Jl{Phys. Lett. B}{#1}}
\def\PRD#1 {\Jl{Phys. Rev. D}{#1}}
\def\PRL#1 {\Jl{Phys. Rev. Lett.}{#1}}

\def\lal{&&\nqq {}}

\def\beq{\begin{equation}}
\def\eeq{\end{equation}}
\def\bear{\begin{eqnarray}}
\def\bearr{\begin{eqnarray} \lal}
\def\ear{\end{eqnarray}}
\def\earn{\nonumber \end{eqnarray}}

\def\nnn{\nonumber\\ \lal }

\begin{document}
\twocolumn[
\jnumber{issue}{year}

\Title{Ruling out an inflation driven by a power law potential: kinetic coupling does not help}

\Aunames{Avdeev N.A.,\au{a,b,1} Toporensky A. V.,\au{b,c,2} }

\Addresses{
\adr a {Department of Astrophysics and Stellar Astronomy, Faculty of Physics, Lomonosov Moscow State University, Leninskie Gory, 1/2, Moscow 119991, Russia}
\adr b {Sternberg Astronomical Institute, Lomonosov Moscow State University, Universitetsky Prospekt, 13, Moscow 119991, Russia}
\adr c {Kazan Federal University, Kazan 420008, Republic of Tatarstan, Russia}
}


\Abstract
	{We demonstrate  that the latest   constraints  on inflationary observables, namely the tensor-to-scalar ratio $r$ and the scalar spectral index $n_{_S}$, from the  Cosmic Background Radiation (CMB) observations are already strong  enough to rule out the model of a scalar field with a power law potential even in the presence of 
kinetic coupling to gravity  with a positive coupling constant. The case for a negative coupling constant needs  special treatment.}
\medskip

] 
\email 1 {NAAvdeev1995@mail.ru}
\email 2 {atopor@rambler.ru}

\section{Introduction}

In the recent years, a plethora of  modified gravity theories have been put forward in order to source cosmic inflation, an epoch of accelerated expansion in the early universe \cite{Starobinsky1,Guth1,Linde1,Albrecht,Linde2,LiLy,Baumann,Mukhanov,Hawking,Starobinsky2} and to match the  inflationary predictions with the observational data \cite{C,Shinji,Planck,BICEP:2021xfz, Futamase,Barrow,Bezrukov,Picon,Ford,Golovnev,Kawasaki,Kallosh,Kawai,Kallosh2,Satoh,Baumann2,Silverstein,Easson,Akrami,Deffayet}. Scalar-tensor theories represent a large set of
of the gravitational theories studied so far, and some class of the theories based on other principles, such as the  quadratic gravity \cite{Salvio}, can also be reformulated in the scalar-tensor form. 
The general class of scalar-tensor theories leading to the second-order differential equations of motion (the so-called Horndeski theory\cite{Horndeski}) is rather wide, depends upon five functions, not fixed by a theory. Hence a thorough  analysis  of  the particular cases of the Horndeski theory might be well-motivated.

Current CMB bounds on the tensor-to-scalar ratio $r$ and the scalar spectral index $n_{_S}$ coming from the latest PLANCK and BICEP/Keck observations, \cite{Planck,BICEP:2021xfz} are  already strong  enough to rule out a number of inflationary theories. Keeping in mind the large  variety of gravitational theories, ruling out some of them  is no less important than constructing new theories. Based on how  things stand at the present, the case of a  minimally coupled scalar field even with a shallow power law potential is observational disfavoured (see, for example, \cite{Swagat}). In the present paper we show that introduction of the kinetic coupling with  a positive coupling constant does not help.

\section{Basic equations}

  We start our analysis  from the action  
\bearr\label{action}
    S = \int d^4 x \sqrt{-g} \bigl( \frac{R}{16\pi} - \frac{1}{2}[g^{\mu\nu}-
    \nnn
    -\kappa G^{\mu\nu}]\phi_{,\mu}\phi_{,\nu}-V(\phi)\bigr) 
\ear
where $R$ is the scalar curvature, $g_{\mu\nu}$ is metric tensor, $G_{\mu\nu}$ is the Einstein tensor, $V(\phi)$ is the scalar field potential which we choose in the power law form $V=V_0 \phi^{\alpha} $, $\kappa$ is the coupling parameter which we consider positive in the present paper. This particular model have been proposed in \cite{G1} and have been studied further in many papers (see, for example,
\cite{G2,G3,G4,G5}).

Field equations of this theory with Fridman-Robertson-Walker metric are as follows
\begin{equation}\label{equations_1}
    3H^{2} = 4\pi\dot \phi^{2}(1+9 \kappa H^{2}) + 8\pi V_0 \phi^{\alpha} \\
\end{equation}  

\bearr\label{equations_2}
 2\dot H + 3 H^{2} = -4 \pi \dot \phi^{2} (1 - \kappa (2 \dot H + 3 H^{2} + 
 \nnn
 +4 H\ddot \phi \dot{\phi}^{-1})) + 8\pi V_0 \phi^{\alpha} 
\ear

\bearr\label{equations_3}
    (\ddot\phi + 3 H \dot\phi) + 3\kappa(H^{2}\ddot\phi + 2 H \dot H \dot \phi + 
     \nnn
    + 3 H^{3} \dot \phi ) = -V_0 \alpha \phi^{\alpha-1}
\ear

During inflation, the scalar field is in a slow-roll phase, so we introduce the slow-roll parameters: 
\bearr\label{slowroll_parameters}
    \epsilon_0 = -\frac{\dot H}{H^2}
    \nnn
    \epsilon_1 = \frac{\dot \epsilon_0}{H \epsilon_0}
    \nnn
    k_0 = 12\pi\kappa\dot\phi^2
    \nnn
    k_1 = \frac{\dot k_0}{H k_0}
\ear
During  slow-roll inflation, we have   $\epsilon_0, \epsilon_1, k_0, k_1<<1$.

The field equations (\ref{equations_1}-\ref{equations_3}) in the slow-roll approximation can be rewritten in the following form:

\bearr\label{equations_slowroll}
    \frac{3H^2}{8\pi} =V_0\phi^{\alpha}
    \nnn
    \dot{H} = -4\pi\dot{\phi^2}-12\pi\kappa H^2 \dot{\phi^2}
    \nnn
    3H\dot{\phi}+\alpha V_0 \phi^{\alpha-1} + 9H^3\kappa\dot{\phi} = 0
\ear
or in the form, explicitly  resolved with respect to highest derivative terms:
\bearr\label{explicit_equations_slowroll}
    \dot{\phi} = -\frac{\alpha V_0 \phi^{\alpha-1}}{2\sqrt{6\pi V_0}\phi^{\alpha/2}(1+8\pi\kappa V_0 \phi^{\alpha})}
    \nnn
    \dot{H} = -\frac{\alpha^2 V_0 \phi^{\alpha-2}}{6(8\pi\kappa V_0\phi^{\alpha}+1)}
    \nnn
    \frac{3H^2}{8\pi} =V_0\phi^{\alpha}
\ear

We can express slow-roll parameters (\ref{slowroll_parameters}) using equations (\ref{explicit_equations_slowroll}):
\bearr\label{slowroll_parameters_phi}
    \epsilon_0 = \frac{\alpha^2}{16\pi\phi^2(1+8\pi\kappa V_0\phi^{\alpha})}
    \nnn
    \epsilon_1 = \frac{\alpha(1+8\pi\kappa V_0\phi^{\alpha}+4\pi\alpha\kappa V_0\phi^{\alpha})}{4\pi\phi^2(1+8\pi\kappa V_0\phi^{\alpha})^2}
\ear
The condition for the end of  inflation is $\epsilon_0 (\phi_E) = 1$, implying
\bearr\label{condition_end}
    \frac{\alpha^2}{16\pi\phi_E^2(1+8\pi\kappa V_0\phi_E^{\alpha})} = 1
\ear
Number of e-folds can be expressed in this form
\bearr\label{efolds}
    N = \int_{\phi_E}^{\phi_I}\frac{H}{\dot\phi}d\phi = -\frac{4\pi}{\alpha}(\phi^2_E-\phi^2_I)-
    \nnn
    -\frac{64\pi^2\kappa V_0}{\alpha(\alpha + 2)}(\phi^{\alpha+2}_E-\phi^{\alpha+2}_I)
\ear
This expression can be used to find initial value of scalar field $\phi_I$. Note, that both $\phi_E$ and $\phi_I$
depend only on the product $\kappa V_0$, we will use this fact later.

\section{Power spectrum}
For the theory under consideration the expressions for the  tensor-to-scalar ratio $r$, scalar spectral index $n_{_S}$ and amplitude of scalar power spectrum $P_{\zeta}$ have been  derived in\cite{C,Shinji} to be 
\bearr\label{ratio}
    r = 16\epsilon_0 = \frac{\alpha^2}{\pi\phi_I^2(1+8\pi\kappa V_0 \phi_I^{\alpha})}~,
\ear

\bearr\label{spectral_index}
    n_{_S} = 1 - 2\epsilon_0 - \epsilon_1 = 1 - \frac{\alpha^2}{8\pi\phi_I^2(1+8\pi\kappa V_0\phi_I^{\alpha})} -
    \nnn
    -\frac{\alpha(1+8\pi\kappa V_0\phi_I^{\alpha}+4\pi\alpha\kappa V_0\phi^{\alpha})}{4\pi\phi_I^2(1+8\pi\kappa V_0\phi_I^{\alpha})^2}
\ear
Note, that $r$ and $n_{_S}$ do not explicitly depend on $k_0$ and $k_1$, it is known that these slow-roll parameters only appears in second order terms \cite{C}. The  role of non-minimal coupling in the slow-roll approximation is to modify equations for standard slow-roll parameters  $\epsilon_0$ and $\epsilon_1$.
Expression for the amplitude of scalar power spectrum  in case of small slow-roll parameters $\epsilon_0, \epsilon_1, k_0, k_1$ takes the following form (taking into account only the leading order term)
\bearr\label{scalar_amplitude}
    P_{\zeta} \approx \frac{H^2}{8\pi^2} \frac{1}{\epsilon_0} = \frac{16 V_0 \phi_I^{\alpha+2}(1+8\pi \kappa V_0 \phi_I^{\alpha})}{3\alpha^2}
\ear
It is important to note that expressions for $r$ and $n_{_S}$ parameters depend only on the product  $\kappa V_0$, but do not depend on them separately. This is the reason why we can  test this theory (and actually rule it out) using PLANCK+BICEP/Keck results \cite{BICEP:2021xfz} by changing the values of $\kappa V_0$. On the contrary, scalar power spectrum amplitude depends on $\kappa$ and $V_0$ separately (\ref{scalar_amplitude}), so it be satisfied easily  by changing these parameters separately.

\begin{figure*}
\centering
\includegraphics[width=5cm]{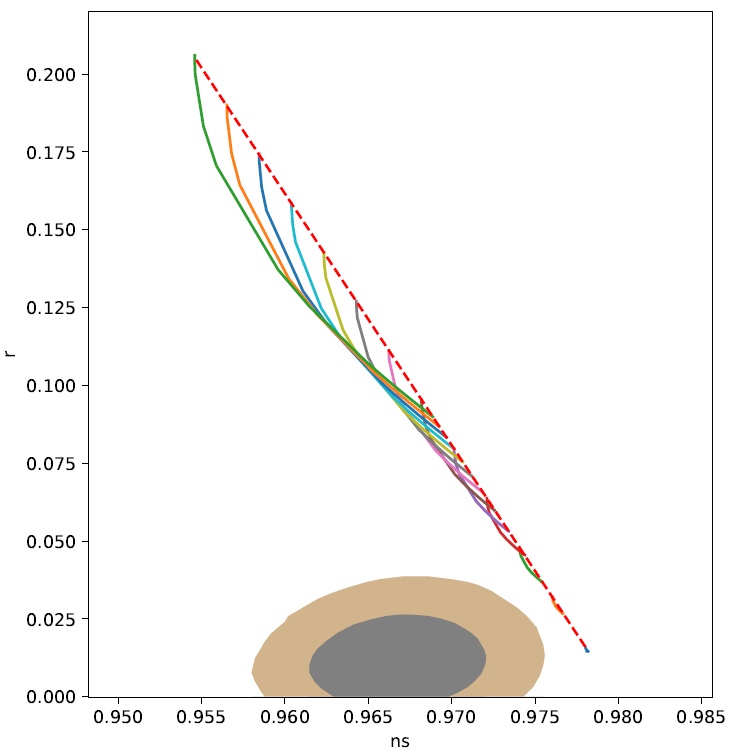} \quad
\includegraphics[width=5cm]{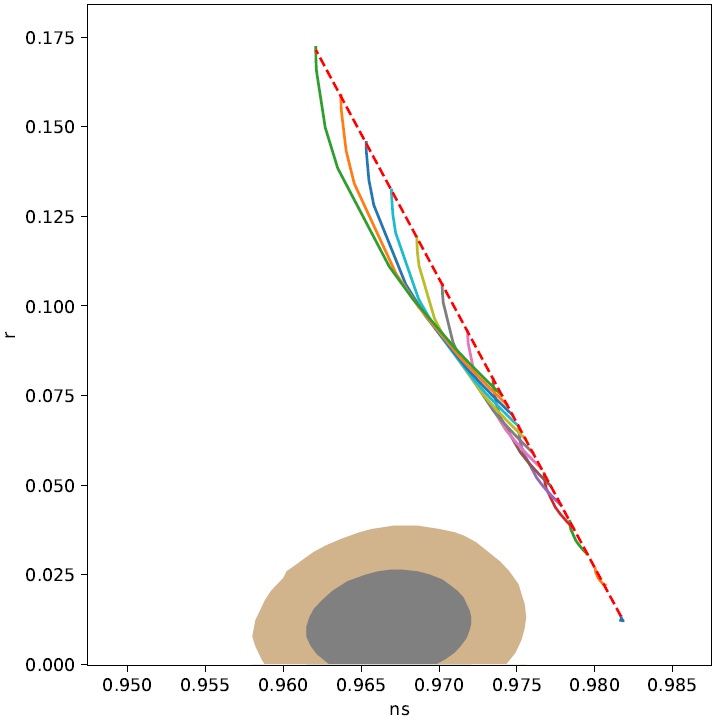} \quad
\includegraphics[width=5cm]{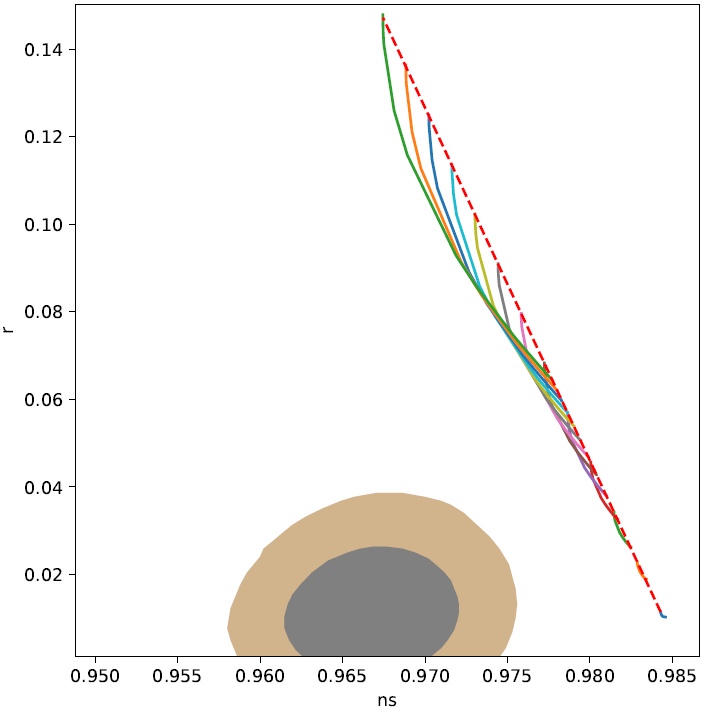}\\
\inch (a) \hspace{4.5cm} (b)\inch \hspace{2.1cm} (c)\inch
\caption{\small 
		In this figure, graphs are plotted in the range of the degree of potential $\alpha = 0.01..2.4$, for  a) N=50, b) N = 60, c) N = 70, each graph is built in a range of values $\kappa V_0 = -0.001..1000$, tan and gray areas describes 68\% and 95\% confidentional levels of Planck results (TT,TE,EE+lowE+lensing+BK15+BAO)\cite{BICEP:2021xfz}, dashed red line describes results for theory with minimal kinetic coupling}
		\label{fig1}
\end{figure*}

We use formulas (\ref{ratio}) and (\ref{spectral_index}) for plotting the dependence  $(r, n_s)$ ( Fig.\ref{fig1}).

As we see adding nonminimal kinetic coupling make results better, but still not good enough to pass the latest test results. An interesting feature of these results is the fact that when $\kappa V_0 \rightarrow \infty$, the curve $(r, n_s)$  ends on a line corresponding to the case $\kappa V_0 = 0$ (of course, for some different $\alpha$). Analytic expressions for case $\kappa V_0 \rightarrow \infty$ can be written in the following form
\bearr\label{limits}
    \lim\limits_{\kappa V_0\to\infty} r = \frac{16\alpha}{2N(\alpha+2)+\alpha} \nnn
    \lim\limits_{\kappa V_0\to\infty} n_s = 1 - \frac{4(\alpha+1)}{2N(\alpha+2)+\alpha}
\ear
We can  now explicitly express the dependence of $n_{_S} - 1$ upon $r$
\bearr\label{assimptotic_dependence}
    n_{_S} - 1 = - \frac{(2N-1)r+16}{16N}
\ear
Analytic expressions for $r$ and $n_{_S}$ for case $\kappa V_0 = 0$  (a minimally coupled scalar field) take the form
\bearr\label{analitic}
    r = \frac{4\alpha}{\frac{\alpha}{4} + N}
    \nnn
    n_{_S} = 1 - \frac{\alpha + 2}{2N+\frac{\alpha}{2}}
\ear
which leads to
\bearr\label{analitic_dependence}
    n_{_S} - 1 = - \frac{(2N-1)r+16}{16N}
\ear

As can be seen from the equations (\ref{assimptotic_dependence}) and (\ref{analitic_dependence}), the line for the limit $\kappa V_0 \to \infty$ coincides with the line for a minimally coupled scalar field. This fact tells us that increasing the $\kappa V_0$ parameter to high values can not improve the result
for a minimally coupled scalar field, which is already ruled out. In the range of intermediate $\kappa V_0$ the $(r, n_s)$ curve deviates from the line of minimally coupled field, however, this deviation is not large  enough  to match the current
observational data. This means that the model with non-minimally kinetically coupled scalar field solely  can be already ruled out for any power law potential.

\section{Conclusion}

In this paper, the scalar-tensor theory with non-minimal kinetic coupling in the case of positive 
coupling constant $\kappa$ and a power law scalar field potential have been considered. Using the latest observational data, it was shown that in this case this theory can be ruled out.

Note that the dynamics of negative $\kappa$ is much richer and includes, for example, such an exotic possibility as realising  inflation without a scalar field
potential \cite{Sushkov}. This regime requires large $\dot \phi$ \cite{Avdeev}, and hence, can not be described by the  usual slow-roll approximations. We leave the case of negative $\kappa$ for a future work.

\Acknow{Authors are grateful to Swagat Saurav Mishra for discussions.}

\Funding{The work of AT and NA have been supported by the RFBR grant 20-02-00411. NA thanks the Foundation for the Theoretical Physics and Mathematics Advancement Foundation “BASIS,” of which he is a fellow(Stipend - Physics faculty (PhD Student): grant 21-2-2-46-1). AT thanks  the Russian Government Program of Competitive Growth of Kazan Federal University.}


\end{document}